\documentclass[10pt,journal,twoside,web]{ieeecolor}
\usepackage{tmi2}
\usepackage{cite}
\usepackage{amsmath,amssymb,amsfonts}
\usepackage{algorithm}
\usepackage{algorithmic}
\usepackage{graphicx}
\usepackage{textcomp}
\def\BibTeX{{\rm B\kern-.05em{\sc i\kern-.025em b}\kern-.08em
    T\kern-.1667em\lower.7ex\hbox{E}\kern-.125emX}}
\usepackage{comment}
\usepackage{subfig}
\usepackage{tikz}
\usetikzlibrary{matrix,calc}
\usepackage{amsmath}
\usepackage{framed,multirow}
\usepackage{amssymb}
\usepackage{latexsym}
\usepackage{url}
\usepackage{xcolor}
\usepackage{lipsum}
\usepackage{hyperref}
\hypersetup{
    colorlinks=true,
    linkcolor=blue,
    filecolor=magenta,      
    urlcolor=cyan,
    pdftitle={DriftRecon},
    }
\usepackage{graphics}
\usepackage{gensymb}
\usepackage{caption}

\usepackage{soul}
\usepackage{amsfonts}
\usepackage{algorithmic}
\usepackage{graphicx}
\usepackage{textcomp}
\usepackage{booktabs}

\usepackage{bbding}

\usepackage{xcolor}
\definecolor{customblue}{RGB}{0, 153, 204}

\definecolor{newcolor}{rgb}{.8,.349,.1}
\def\baselinestretch{1.0}
\makeatletter
\IEEEtriggercmd{\reset@font\normalfont\fontsize{7.9pt}{8.40pt}\selectfont}
\makeatother
\IEEEtriggeratref{1}

\begin{document}
\title{Guidance-Coupled Nullspace Drifting for\\ One-Step Medical Image Reconstruction}
\author{M. Tuna Alatan, A. Server Tapan, Saban Ozturk, M. Usama Mirza, Bilal Kabas, Ge Wang, and Tolga \c{C}ukur \vspace{-1.2cm}
\\
\thanks{This work was supported by TUBA GEBIP 2015 and BAGEP 2017 fellowships awarded to T. \c{C}ukur. (Corresponding authors: S. Ozturk, saban.ozturk@hbv.edu.tr and T. \c{C}ukur, cukur@ee.bilkent.edu.tr).}
\thanks{M.T. Alatan, A.S. Tapan, M. U. Mirza, B. Kabas, and T. \c{C}ukur are with the Department of Electrical and Electronics Engineering and the National Magnetic Resonance Research Center (UMRAM), Bilkent University, Ankara, Turkey. S. Ozturk is with UMRAM, Bilkent University, Ankara, Turkey. T. \c{C}ukur is also with the Neuroscience Program, Bilkent University, Ankara, Turkey. G. Wang is with the Biomedical Imaging Center, Rensselaer Polytechnic Institute, New York, USA.}
}

\maketitle

\begin{abstract}
Generative reconstruction can recover high-quality images from undersampled acquisitions, but state-of-the-art generative priors commonly rely on iterative sampling. Reducing the number of iterations lowers inference cost but also removes opportunities for progressive distributional refinement, making it difficult to retain image realism without sacrificing reconstruction fidelity. Drift modeling offers an alternative by shifting progressive refinement to training, thereby enabling one-step sampling without an iterative inference trajectory. Existing drifting methods, however, do not account for the imaging operator when modeling distributional drift. We propose DriftRecon, a generative method for one-step medical image reconstruction that performs drifting directly in the nullspace of the imaging operator. DriftRecon uses nullspace-restricted noise to construct training samples and a multiscale generator that incorporates measurement conditioning across resolution levels, with its generated component projected onto the operator nullspace. To control the fidelity--realism trade-off at inference, a conditional guidance parameter continuously steers the same generator between measurement-conditioned and unconditional generation. We derive a guidance-dependent similarity-kernel temperature schedule that preserves the guidance-weighted mass receiving strengthened supervision, thereby progressively localizing drift interactions as guidance increases to limit the associated loss of reconstruction fidelity. Experiments in sparse-view CT and accelerated MRI reconstruction show that DriftRecon outperforms established generative reconstruction methods and recent drifting models while producing reconstructions in a single forward pass.
\vspace{-0.2cm}
\end{abstract}

\begin{IEEEkeywords}
medical image reconstruction, nullspace, drifting, MRI, CT \vspace{-5mm}
\end{IEEEkeywords}

\bstctlcite{IEEEexample:BSTcontrol}

\section{Introduction}
Medical image reconstruction aims to recover high-quality images from undersampled data acquired according to a prescribed imaging operator determined by the acquisition physics and sampling pattern \cite{MIR-survey1,mrireview}. Because reconstruction is an underdetermined problem, reliable recovery depends jointly on physical guidance from the imaging operator and learned image priors that capture the anatomical structure and detailed characteristics of high-quality images \cite{Schlemper2017,Hammernik2017,DreamNet,bindong2021,MoDl}. In recent years, generative image priors based on diffusion and flow modeling have made substantial progress in this setting, recovering high-quality images from accelerated MRI \cite{peng2022towards,huang2023cdiffmr,fdb,luo2025upmri} and sparse-view CT \cite{Chung_2023_CVPR,CheJix_Crossview_MICCAI2025,li2026cdpir,force_tmi} protocols. By modeling a distribution of plausible images rather than a single deterministic mapping, generative methods can also support different reconstruction characteristics along the fidelity--realism trade-off \cite{fabian2024diracdiffusion,bendel2024pcagan}. Despite these advantages, these methods commonly rely on iterative sampling processes that map noise or undersampled data toward reconstructions consistent with fully-sampled acquisitions through a sequence of smaller mappings \cite{peng2022towards,fdb,Chung_2023_CVPR,luo2025upmri}. Through successive updates, iterative sampling progressively refines the reconstruction toward the target image distribution while providing repeated opportunities to correct errors in anatomical structure. The resulting inference process, however, incurs substantial computational cost, which limits the practical utility of generative reconstruction.

Since reducing generative reconstruction to a single network evaluation could improve inference efficiency while, in principle, retaining the distributional benefits of learned generative priors, several acceleration strategies have been proposed in the literature. Trajectory-compression methods include implicit sampling that skips intermediate steps \cite{song2021ddim,chung2022cvpr}, progressive distillation that transfers a multistep teacher sampler into successively shorter student samplers \cite{salimans2022progressive}, and consistency learning that trains predictions from different stages of the sampling process to approach a common solution \cite{song2023consistency,akoda2026cmore}. Task-specific trajectory design instead constructs the generative process around the reconstruction problem so that shorter sampling trajectories are supported inherently \cite{ssdiffrecon,CheJix_Crossview_MICCAI2025}. Regardless of the strategy, compressing sampling toward a single evaluation removes the successive updates through which generative reconstruction progressively refines the image distribution \cite{kazerouni2023diffusion}. One-step generation instead has to recover measurement-consistent image structure and realistic unresolved detail within a single mapping, without subsequent opportunities for correction \cite{gungor2022adaptive}. The central challenge is therefore to retain the realism afforded by generative reconstruction without sacrificing reconstruction fidelity under one-step inference.

Drift modeling offers an emerging alternative by shifting progressive distribution refinement from inference to training, during which the generator distribution is iteratively moved toward the target image distribution \cite{deng2026drifting}. The refinement normally accumulated over multiple inference steps is therefore learned during training and realized in a single generator evaluation. This training-time evolution is driven by an attraction--repulsion field constructed between synthetic and target samples, with similarity-kernel temperatures controlling the locality of the corresponding interactions. Recent medical imaging studies have begun to explore conditional drifting with promising results across low-dose CT denoising \cite{wang2026rddm}, CT image synthesis \cite{li2026gdm}, and MRI reconstruction \cite{lyu2026sardmdc}. Despite these encouraging results, current approaches do not incorporate the imaging operator into the modeled distributional drift. In \cite{wang2026rddm,li2026gdm} the drift is learned without reference to the underlying imaging operator, whereas \cite{lyu2026sardmdc} performs residual drifting before posthoc data-consistency (DC) enforcement. The drift can therefore extend into the range space of the imaging operator, causing the generator to modify image components already determined by the acquired data, so part of the generative update is subsequently removed without contributing to the final reconstruction \cite{chung2022score,schwab2019deep}. Existing drifting methods also offer limited control over the fidelity--realism trade-off: separate models with different similarity-kernel temperatures provide discrete operating points \cite{wang2026rddm}, whereas conditional--unconditional guidance enables continuous control within a single model but has been used with a fixed kernel temperature \cite{deng2026drifting}. Because recent theoretical analysis suggests that guidance strength and kernel temperature jointly determine the conditional drifting field \cite{lai2026unified}, we hypothesize that increasing guidance without adapting the kernel temperature can strengthen mismatched drift interactions and thereby reduce reconstruction fidelity.

\begin{figure}[t]
        \vspace{0.1cm}
        \centerline{\includegraphics[width=0.975\columnwidth]{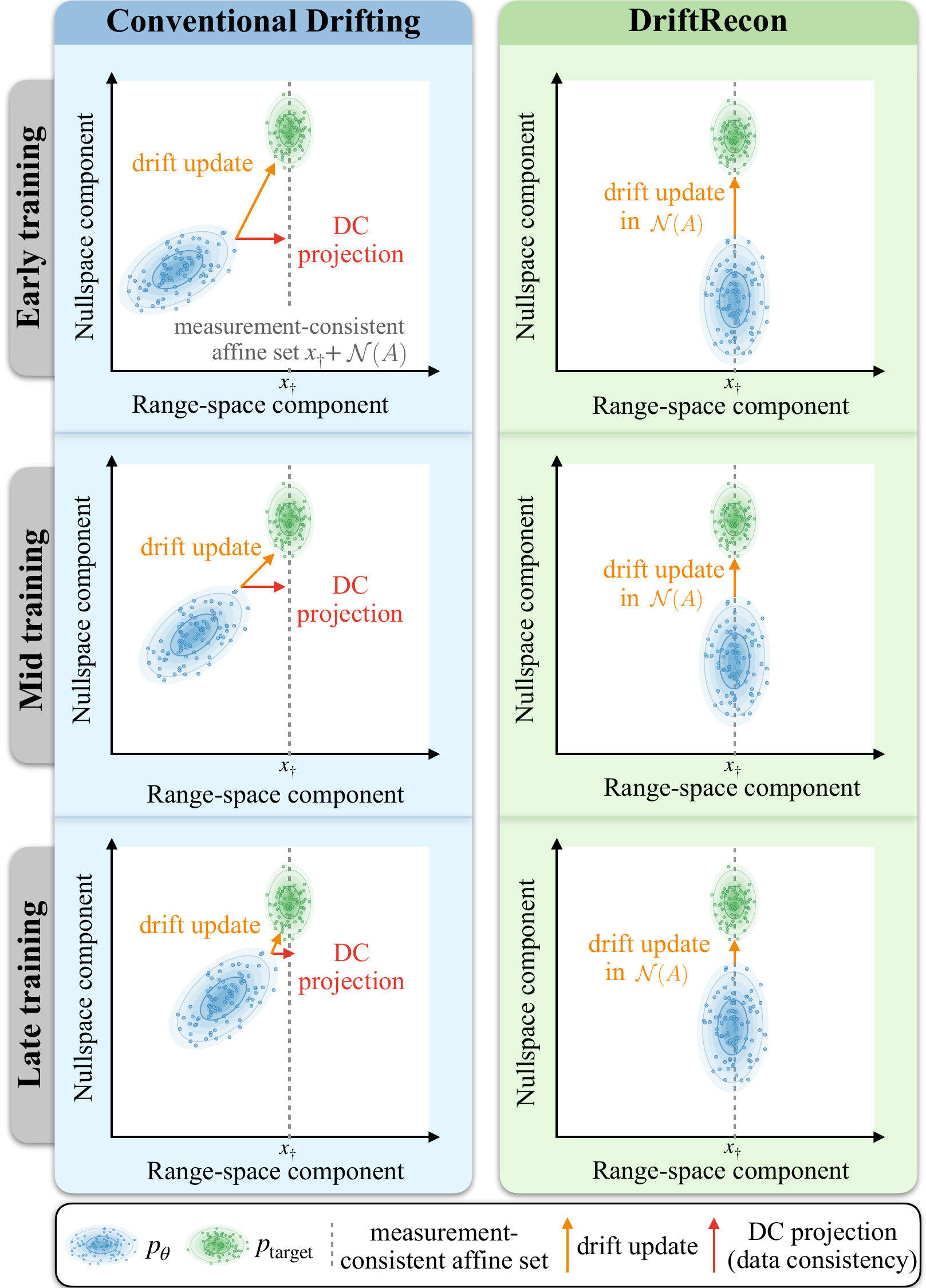}}
        \caption{\textbf{Operator-nullspace drifting.} Conventional drifting can alter measurement-determined components, requiring data-consistency projection onto $x_{\dagger}+\mathcal{N}(\mathcal{A})$. DriftRecon confines generation and drift to $\mathcal{N}(\mathcal{A})$, preserving the fixed anchor $x_{\dagger}=\mathcal{A}^{\dagger}y$ and ensuring $\mathcal{A}\hat x=y$ throughout training. Blue and green distributions denote the generated distribution $p_{\theta}$ and target distribution $p_{\mathrm{target}}$, respectively; orange arrows indicate drift updates, and red arrows indicate data-consistency projections.}
        \label{fig:nullspace}
\end{figure}

\begin{figure*}[t]
        \vspace{0.1cm}
        \centerline{\includegraphics[width=0.95\textwidth]{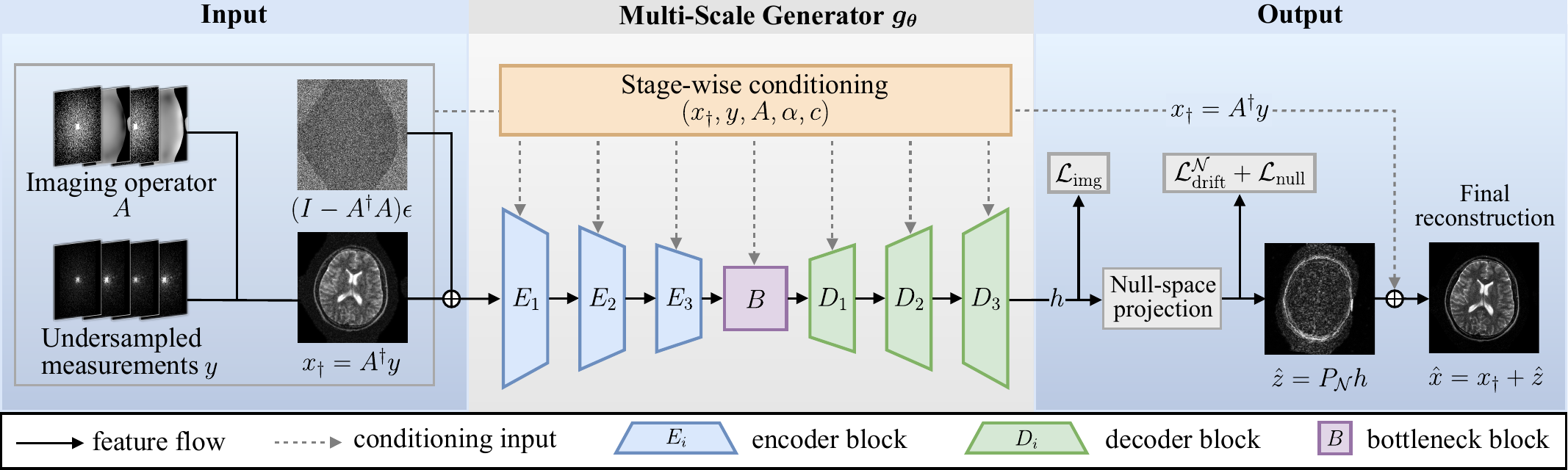}}
        \caption{\textbf{Generator architecture.} The multiscale generator \(g_{\theta}\) receives the nullspace-projected noise \(\epsilon^{\mathcal N}\) and measurement-consistent reconstruction \(x_{\mathrm{\dagger}}\), together with the imaging operator \(\mathcal A\), measurements \(y\), and guidance parameter \(\alpha\). Measurement conditioning is incorporated across network stages as the inputs are processed through the encoder and decoder. The resulting drift estimate \(h\) is projected onto the operator nullspace as \(\hat{z}=\mathcal P_{\mathcal N}h\) and added to \(x_{\mathrm{\dagger}}\) to yield the final reconstruction \(\hat x=x_{\mathrm{\dagger}}+ \hat{z}\) in a single forward pass.}
        \label{fig:architecture}
\end{figure*}

Here we propose DriftRecon, a generative method for one-step medical image reconstruction that models distributional drift directly in the nullspace of the imaging operator. By restricting the drift to components not determined by the acquired data, DriftRecon preserves measurement-supported information while focusing generative modeling on the unresolved degrees of freedom of the reconstruction problem (Fig. \ref{fig:nullspace}). This is realized through nullspace-restricted noise for constructing training samples and a multiscale generator that incorporates measurement conditioning across resolution levels before projecting its generated component onto the operator nullspace (Fig. \ref{fig:architecture}). The imaging operator therefore constrains the distributional drift itself rather than being applied posthoc. To provide continuous control over the fidelity--realism trade-off, DriftRecon steers its generator between measurement-conditioned and unconditional generation. Increasing guidance, however, also strengthens the influence of mismatched real samples on the drift, which can compromise reconstruction fidelity when similarity-kernel temperature remains fixed. We therefore derive a guidance-dependent temperature schedule by preserving the guidance-weighted mass receiving strengthened supervision, progressively localizing drift interactions as guidance increases (Fig. \ref{fig:temperature}). Comprehensive experiments on sparse-view CT and accelerated MRI demonstrate the effectiveness of DriftRecon against data-driven and physics-based baselines as well as diffusion, flow, and recent drifting approaches, while enabling continuous fidelity--realism control from a single one-step model. Code is available at {\small \url{https://github.com/icon-lab/DriftRecon}}.

\vspace{0.1cm}
\subsubsection*{\textbf{Contributions}}
\begin{itemize}
\item We introduce operator-nullspace drifting as a generative prior for medical image reconstruction, restricting learned image generation to degrees of freedom unresolved by the acquired measurements while preserving measurement-supported information.

\item We design a one-step drift model combining nullspace-restricted training perturbations with a multiscale generator that incorporates measurement conditioning across resolution levels and projects its generated component onto the operator nullspace.

\item We derive a guidance-dependent drift temperature schedule by preserving the guidance-weighted mass receiving strengthened supervision, progressively localizing drift interactions as guidance increases to maintain reconstruction fidelity.
\end{itemize}

\section{Related Work}

\subsection{Efficient Generative Reconstruction}
Recent generative reconstruction methods increasingly seek to reduce the computational cost of iterative diffusion and flow sampling \cite{peng2022towards,huang2023cdiffmr,Chung_2023_CVPR,luo2025upmri,force_tmi}. Implicit sampling \cite{song2021ddim,chung2022cvpr}, progressive distillation \cite{salimans2022progressive}, and reconstruction-specific trajectory design \cite{ssdiffrecon,CheJix_Crossview_MICCAI2025,fdb} can substantially shorten the sampling process, often to the order of tens of network evaluations, but inference remains iterative. Compressing these processes further toward a single evaluation removes the successive updates through which the image distribution is progressively refined, making it increasingly difficult to retain realism while preserving reconstruction fidelity. Consistency learning directly targets this regime by training predictions from different stages of the sampling process to approach a common solution \cite{song2023consistency,akoda2026cmore}. Although this enables one-step generation, medical image reconstruction with a consistency prior can still require iterative conjugate-gradient data-consistency updates after the network evaluation, partially offsetting the efficiency benefit of one-step inference \cite{akoda2026cmore}. Drift modeling takes a different route by moving the progressive evolution of the generator distribution to training, allowing the learned reconstruction mapping itself to be executed in a single network evaluation \cite{deng2026drifting}.

\subsection{Drift Modeling in Medical Imaging}
Several recent studies have considered drifting for image synthesis and reconstruction tasks. \cite{wang2026rddm} formulates residual drifting between low- and normal-dose CT distributions. \cite{li2026gdm} constructs a conditional drift between source and target image distributions for CT image synthesis and reconstruction. \cite{lyu2026sardmdc} adopts residual drifting for MRI reconstruction, defining the drift over a reconstruction residual before posthoc DC enforcement. These recent methods do not impose an explicit constraint that restricts the learned drifting process to the operator nullspace. Consequently, the resulting drift estimates can contain components orthogonal to the nullspace that conflict with the acquired data. DriftRecon instead constrains both training perturbations and predicted drift estimates to the operator nullspace, confining the learned distributional drift to degrees of freedom not determined by measurements.

A second challenge is controlling the fidelity--realism trade-off without sacrificing reconstruction fidelity. Existing drifting formulations approach this in two ways. One uses separately configured models with different similarity-kernel temperatures to obtain distinct fidelity--realism balances \cite{wang2026rddm}, providing only discrete control and requiring multiple trained models. The other uses conditional--unconditional guidance to vary this balance continuously within a single model \cite{deng2026drifting}, but retains a fixed kernel temperature across guidance strengths. Because kernel temperature determines the locality of the attraction--repulsion interactions, stronger guidance can disrupt the conditional--unconditional balance without adapting the locality of the drift interactions \cite{lai2026unified}, which can degrade reconstruction fidelity. DriftRecon instead decreases kernel temperature with increasing guidance, progressively localizing drift interactions to limit fidelity loss while retaining continuous control over the fidelity--realism trade-off.

\section{Theory}

\subsection{Conditional Drifting for Reconstruction}

Medical image reconstruction entails solving the following linear inverse problem:
\begin{equation}
\label{eq:forward_model}
y=\mathcal{A}x^{\star},
\end{equation}
where $\mathcal{A}$ is the imaging operator, $y$ denotes the undersampled measurements, and $x^{\star}$ is the target image. Conventional drifting formulations can use the least-squares reconstruction $x_{\dagger}=\mathcal{A}^{\dagger}y$ as the conditioning image, where $\mathcal{A}^{\dagger}$ is the Moore-Penrose pseudoinverse. Given Gaussian noise $\epsilon\sim p_{\epsilon}$ and a scalar guidance parameter $\alpha\geq1$, the generator $f_{\theta}$ produces:
\begin{equation}
\label{eq:conventional_generator}
\hat x=f_{\theta}(\epsilon,x_{\dagger},\alpha), \,\,\, q_{\theta}(\cdot\mid x_{\dagger},\alpha)=\big[f_{\theta}(\cdot,x_{\dagger},\alpha)\big]_{\#}p_{\epsilon},
\end{equation}
where $q_{\theta}$ is the distribution of images produced for different noise samples under the same condition, and $[\cdot]_{\#}$ denotes the pushforward of a distribution. Training then requires positive and negative samples that define the distributional drift, a similarity kernel that weights their interactions, and an objective that updates the generator toward the resulting attraction--repulsion field.

For a given $\hat x$, let $x^{+}$ denote the target image matching $x_{\dagger}$ (i.e., the ground-truth image derived from the fully-sampled data paired with the current undersampled acquisition). Generated images from $f_{\theta}$ under the same condition form negative samples $x^{-}$, while additional real images $x_{\mathrm{mis}}$ that do not match $x_{\dagger}$ (e.g., ground-truth images from a different subject) provide mismatched negatives for conditional--unconditional guidance. If $p_{\mathrm{mis}}(\cdot\mid x_{\dagger})$ denotes their distribution, the guided negative distribution for $x^-$ is:
\begin{equation}
\label{eq:guided_negative_distribution}
\widetilde q_{\theta,\alpha}=\frac{1}{\alpha}q_{\theta}(\cdot\mid x_{\dagger},\alpha)+\left(1-\frac{1}{\alpha}\right)p_{\mathrm{mis}}(\cdot\mid x_{\dagger}).
\end{equation}
Hence, $\alpha=1$ uses only generated negatives, whereas higher $\alpha$ increases the contribution of mismatched real samples.

The contribution of individual samples to the distributional drift is determined by the similarity kernel:
\begin{equation}
\label{eq:kernel_generic}
k_{\tau}(x,x')=\exp\left(-\frac{d(x,x')}{\tau}\right),
\end{equation}
where $d(\cdot,\cdot)$ is the kernel distance and $\tau>0$ is the similarity-kernel temperature. Following \cite{deng2026drifting}, the attraction--repulsion field for the generated image $\hat x$ is:
\begin{equation}
\label{eq:conditional_drift}
\mathcal{V}_{\theta}^{(\tau)}(\hat x,\alpha)=
\frac{\mathbb{E}_{x^{+},x^{-}}\!\left[k_{\tau}(\hat x,x^{+})k_{\tau}(\hat x,x^{-})(x^{+}-x^{-})\right]}
{\mathbb{E}_{x^{+}}\!\left[k_{\tau}(\hat x,x^{+})\right]\mathbb{E}_{x^{-}}\!\left[k_{\tau}(\hat x,x^{-})\right]},
\end{equation}
where $x^{+}\sim p_{\mathrm{data}}(\cdot\mid x_{\dagger})$ and $x^{-}\sim\widetilde q_{\theta,\alpha}$. The field attracts generated samples toward matched targets while incorporating repulsion from generated and mismatched negatives. It defines the stop-gradient target used to train the generator:
\begin{equation}
\label{eq:conventional_drift_loss}
\mathcal{L}_{\mathrm{drift}}=
\mathbb{E}_{\epsilon\sim p_{\epsilon},\,\alpha\sim p_{\alpha}}
\!\left[
\left\|
\hat x-\operatorname{sg}\!\left(\hat x+\mathcal{V}_{\theta}^{(\tau)}(\hat x,\alpha)\right)
\right\|_2^2
\right],
\end{equation}
where $\operatorname{sg}(\cdot)$ denotes stop-gradient and $\alpha\sim p_{\alpha}$ controls the balance between measurement-conditioned and unconditional generation \cite{deng2026drifting}. Repeated updates move the generator distribution toward the conditional target distribution. Once trained, reconstruction requires a single evaluation as in Eq.~\eqref{eq:conventional_generator}, but neither the generated image nor the distributional drift is explicitly constrained by $\mathcal{A}$.

\subsection{DriftRecon}
\begin{figure}[t]
        \vspace{0.1cm}
        \centerline{\includegraphics[width=0.99\columnwidth]{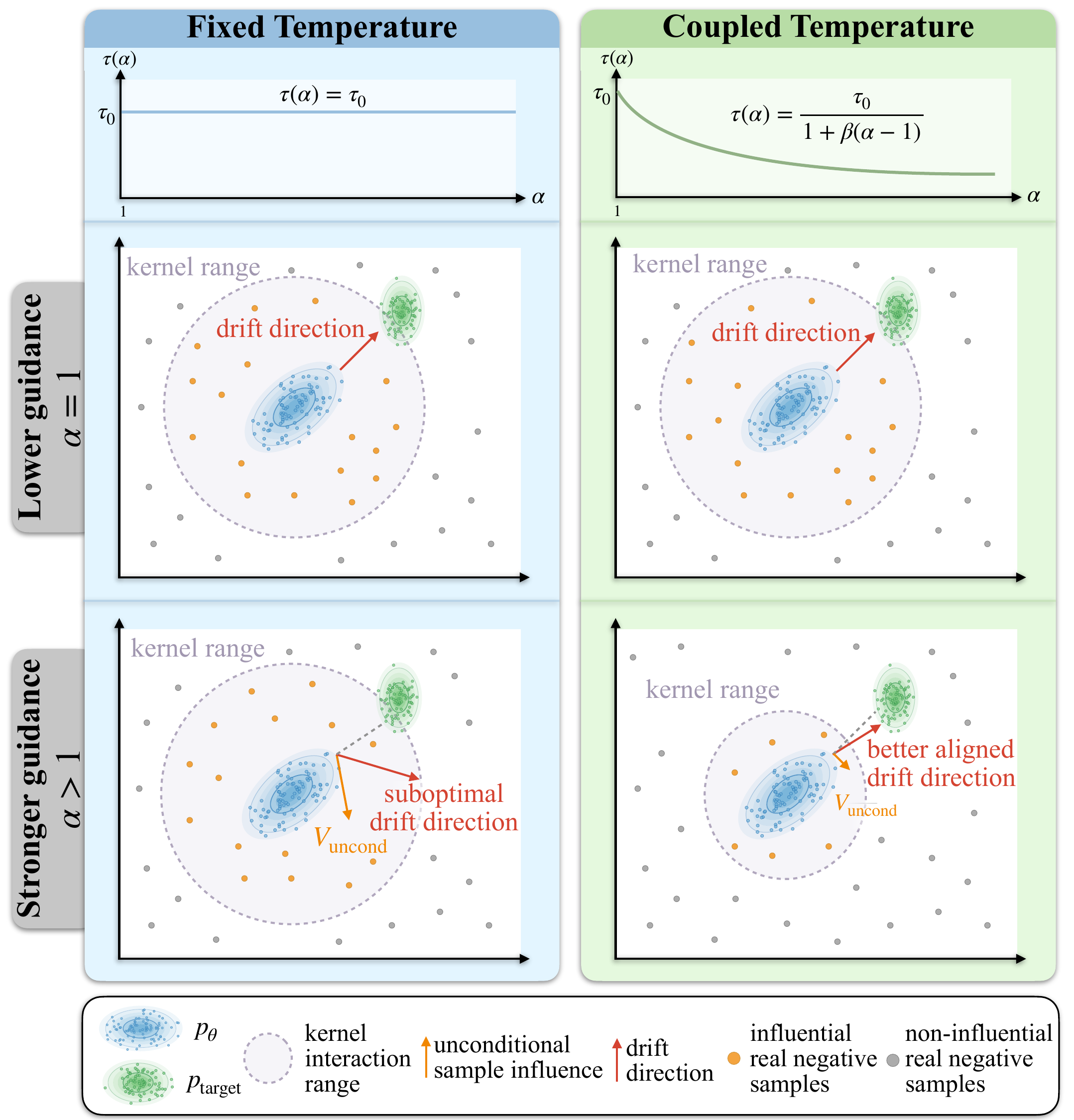}}
        \caption{\textbf{Guidance-coupled drifting.} Fixed-temperature drifting (left) versus the proposed guidance-coupled temperature schedule (right). The top row shows the temperature schedules across guidance strength $\alpha$. At $\alpha=1$ (middle row), both schemes provide equivalent drift supervision. At $\alpha>1$ (bottom row), fixed temperature allows distant mismatched samples to distort the drift direction, whereas decreasing temperature localizes the interactions and keeps supervision directed toward nearby samples from the target distribution.}
        \label{fig:temperature}
\end{figure}

DriftRecon introduces two complementary mechanisms for one-step reconstruction: operator-nullspace drifting to preserve measurement consistency in generated reconstructions (Fig.~\ref{fig:nullspace}), and a guidance-dependent similarity-kernel temperature schedule derived from a locality-preservation criterion for fidelity--realism control (Fig.~\ref{fig:temperature}).

\subsubsection{Operator-Nullspace Formulation}
An unrestricted distributional drift can modify image components already determined by the acquired measurements, producing updates that conflict with the forward model and compromise reconstruction fidelity unless subsequently corrected. DriftRecon instead confines the learned prior to the unresolved degrees of freedom by formulating both the stochastic perturbation and generated component in the nullspace of the imaging operator.

During minibatch training, let $(y_i,\mathcal{A}_i,x_i^{\star})$ denote the $i$th paired example, with $\alpha_i\sim p_{\alpha}$, define:
\begin{equation}
\label{eq:nullspace_projector}
x_{\dagger,i}=\mathcal{A}_i^{\dagger}y_i;\,\,\,\,\,\, \mathcal{P}_{\mathcal N_i}=I-\mathcal{A}_i^{\dagger}\mathcal{A}_i; \,\,\,\,\,\, \mathcal{A}_i\mathcal{P}_{\mathcal N_i}=0;
\end{equation}
where $I$ is the identity operator, $x_{\dagger,i}$ serves as a fixed anchor of the measurement-consistent affine set, and $\mathcal{P}_{\mathcal N_i}$ is the orthogonal projector onto the nullspace $\mathcal{N}(\mathcal{A}_i)$. Since $x_{\dagger,i}$ lies in the range of $\mathcal{A}_i^{H}$, $\mathcal{P}_{\mathcal N_i}x_{\dagger,i}=0$, and the set of images consistent with $y_i$ can be expressed as \cite{schwab2019deep}:
\begin{equation}
\label{eq:affine_solution_set}
x_{\dagger,i}+\mathcal{N}(\mathcal{A}_i)=\left\{x_{\dagger,i}+z:\;z\in\mathcal{N}(\mathcal{A}_i)\right\}.
\end{equation}
Under the forward model in Eq.~\eqref{eq:forward_model}, the corresponding range-nullspace decomposition of the target image is:
\begin{align}
\label{eq:nullspace_target_derivation}
x_i^{\star}&=x_{\dagger,i}+z_i^{\star}, \quad z_i^{\star}=\mathcal{P}_{\mathcal N_i}x_i^{\star}\in\mathcal{N}(\mathcal{A}_i), \nonumber \\
&  \mbox{ where } \mathcal{A}_i(x_i^{\star}-x_{\dagger,i}) = y_i-\mathcal{A}_i\mathcal{A}_i^{\dagger}y_i=0.
\end{align}
Thus, $x_{\dagger,i}$ contains the component fixed by the measurements, whereas $z_i^{\star}$ is the unresolved component that must be recovered from the learned prior.

DriftRecon restricts both the stochastic perturbation and the generated component to this unresolved subspace (Fig. \ref{fig:architecture}). During training, independent noise samples $\epsilon_{i,r}\sim p_\epsilon$, $r\in\{1,2\}$, are added in nullspace directions as $\epsilon_{i,r}^{\mathcal N}=\sigma \mathcal{P}_{\mathcal{N}_i}\epsilon_{i,r}$, at a fixed noise scale 
$\sigma$. The generator $g_\theta$ then synthesizes two reconstructions while additionally receiving $\mathcal{A}_i$ and $y_i$:
\begin{equation}
\label{eq:recon}
\begin{aligned}
\hat x_{i,r}&=x_{\dagger,i}+\hat z_{i,r},\quad\text{with}\\ h_{i,r}&=g_\theta(\epsilon_{i,r}^{\mathcal N},x_{\dagger,i},\mathcal{A}_i,y_i,\alpha_i), \quad \hat z_{i,r}=\mathcal{P}_{\mathcal{N}_i}h_{i,r}.
\end{aligned}
\end{equation}
Here $h_{i,r}$ denotes the generator output before projection and $\hat z_{i,r}$ its nullspace-projected component. Applying the imaging operator to the reconstructed image shows that the nullspace restriction preserves the acquired measurements:
\begin{equation}
\label{eq:recon_consistency}
\mathcal{A}_i\hat x_{i,r}
=\mathcal{A}_i(x_{\dagger,i}+\mathcal{P}_{\mathcal{N}_i}h_{i,r})=y_i .
\end{equation}

In contrast, applying a conventional unrestricted drift to the same measurement-consistent reconstruction would give:
\begin{equation}
\label{eq:conventional_drift}
\hat x_i^{\prime}=\hat x_i+\eta\mathcal{V}_{\theta}^{(\tau)}(\hat x_i,\alpha_i), \,\,\,
\mathcal{A}_i\hat x_i^{\prime}=y_i+\eta\mathcal{A}_i\mathcal{V}_{\theta}^{(\tau)}(\hat x_i,\alpha_i),
\end{equation}
which does not in general satisfy the acquired measurements as the unrestricted drift contains components outside $\mathcal{N}(\mathcal{A}_i)$.

\subsubsection{Guidance-Coupled Temperature Scheduling}
Increasing the guidance strength $\alpha$ increases the contribution of mismatched real negatives to the conditional drift. At fixed similarity-kernel temperature, this scaling applies regardless of sample distance, so increasingly distant negatives can influence the reconstruction as guidance becomes stronger. Although these updates remain measurement-consistent in our operator-nullspace formulation, they can still move the reconstruction away from its target and reduce reconstruction fidelity.

Here we derive a guidance-dependent temperature by requiring stronger guidance to concentrate its additional influence on progressively closer mismatched negatives rather than increase the total guidance-weighted mass receiving strengthened supervision. For a minibatch of $B$ conditions, guidance assigns each mismatched real negative a relative weight of $(\alpha-1)/(B-1)$. For a generated negative $x_{\mathrm{gen}}$ and mismatched real negative $x_{\mathrm{mis}}$, define their excess distance from the current reconstruction $x$ as $\Delta(x)=d(x,x_{\mathrm{mis}})-d(x,x_{\mathrm{gen}})$. Under a general temperature $\tau(\alpha)$, the relative contribution of the mismatched negative is:
\begin{equation}
\label{eq:relative_guidance_weight}
\rho(\alpha,\Delta)=\frac{\alpha-1}{B-1}\exp\!\left[-\frac{\Delta}{\tau(\alpha)}\right].
\end{equation}
For $\alpha>1$, a given temperature schedule determines an excess-distance boundary $\Delta_c(\alpha)$: mismatched negatives closer than this boundary gain relative influence as guidance increases, whereas those farther away lose relative influence. Differentiating Eq.~\eqref{eq:relative_guidance_weight} gives this schedule-induced boundary:
\begin{equation}
\label{eq:locality_boundary}
\begin{aligned}
\frac{\partial}{\partial\alpha}\log\rho(\alpha,\Delta)
&=\frac{1}{\alpha-1}-\Delta\frac{\partial}{\partial\alpha}\left(\frac{1}{\tau(\alpha)}\right),\\
\Delta_c(\alpha)
&=\frac{1}{(\alpha-1)\displaystyle\frac{\partial}{\partial\alpha}\left(1/\tau(\alpha)\right)}.
\end{aligned}
\end{equation}
Hence, Eq.~\eqref{eq:locality_boundary} tells us where the boundary falls for a given temperature schedule. 

Next, we determine the desired boundary by requiring the additional influence introduced by stronger guidance to be concentrated within a progressively smaller neighborhood. This construction is inspired by the principles of adaptive-neighborhood methods, where the neighborhood scale is adjusted to control the local sample mass \cite{loftsgaarden1965nonparametric}. Let $p_{\Delta}$ denote the density of mismatched negatives over excess distance. The guidance-weighted mass within $\Delta<\Delta_c(\alpha)$ is:
\begin{equation}
\label{eq:local_guidance_mass}
\begin{aligned}
M(\alpha)=\frac{\alpha-1}{B-1}\int_{0}^{\Delta_c(\alpha)}p_{\Delta}(\delta)\,d\delta \approx\frac{\alpha-1}{B-1}p_{\Delta}(0)\Delta_c(\alpha),
\end{aligned}
\end{equation}
where the approximation assumes that $p_{\Delta}$ varies slowly within this local region. Keeping $M(\alpha)$ constant means that the increasing guidance weight is compensated by a shrinking neighborhood, so stronger guidance concentrates supervision on progressively closer samples rather than increasing the total strengthened supervision. We refer to this requirement as local guidance-mass preservation, which yields the bound:
\begin{equation}
\label{eq:locality_criterion}
\Delta_c(\alpha)=\frac{\tau_0}{\beta(\alpha-1)}, \qquad \beta>0,
\end{equation}
where $\tau_0$ is the reference temperature at $\alpha=1$ and constant factors from Eq.~\eqref{eq:local_guidance_mass} are absorbed into $\beta$.

Finally, we set the schedule-induced and desired boundaries described in Eqs. \eqref{eq:locality_boundary}, \eqref{eq:locality_criterion} to be equal for every guidance strength, resulting in a first-order differential equation whose solution gives $\tau(\alpha)$:
\begin{equation}
\label{eq:tau_schedule}
\begin{aligned}
\frac{\partial}{\partial\alpha}\left(\frac{1}{\tau(\alpha)}\right)&=\frac{\beta}{\tau_0} \,\,\,\rightarrow\,\,\, \tau(\alpha)=\frac{\tau_0}{1+\beta(\alpha-1)}.
\end{aligned}
\end{equation}
The limiting case $\beta=0$ recovers fixed-temperature drifting. For $\beta>0$, the derived schedule contracts the region of mismatched negatives whose relative influence is strengthened as guidance increases. Guidance strength therefore controls both the fidelity--realism trade-off and the locality of drift supervision through a single scalar parameter (Fig.~\ref{fig:temperature}).

\subsubsection{Learning the Nullspace Drift Field}
The nullspace formulation and guidance-dependent temperature are combined during training by constructing the attraction--repulsion field from measurement-specific nullspace samples. For condition $i$, the positive sample is
$z_i^\star=\mathcal{P}_{\mathcal{N}_i}x_i^\star$, while the remaining ground-truth images provide mismatched real negatives
$z_{ij}^{\mathrm{mis}}=\mathcal{P}_{\mathcal{N}_i}x_j^\star$, $j\neq i$.
The two generated samples $\hat z_{i,1}$ and $\hat z_{i,2}$ serve as each other's generated negative.

We state the field for $\hat z_{i,1}$; that for $\hat z_{i,2}$ follows by exchanging the subscripts. Let
$k_i^{\mathrm{gen}}=k_{\tau(\alpha_i)}(\hat z_{i,1},\hat z_{i,2})$ and $k_{ij}=k_{\tau(\alpha_i)}(\hat z_{i,1},z_{ij}^{\mathrm{mis}})$.
With $w_i=(\alpha_i-1)/(B-1)$, substituting the guided negative mixture into Eq.~\eqref{eq:conditional_drift} gives:
\begin{equation}
\label{eq:null_drift_field}
\mathcal{V}_{i,1}^{\mathcal N}(\alpha_i) = \frac{ k_i^{\mathrm{gen}}(z_i^\star-\hat z_{i,2}) + w_i\sum_{j\neq i}k_{ij}(z_i^\star-z_{ij}^{\mathrm{mis}})
}{k_i^{\mathrm{gen}} + w_i\sum_{j\neq i}k_{ij} }.
\end{equation}
At $\alpha_i=1$, the mismatched terms vanish and:
\begin{equation}
\label{eq:null_drift_unguided}
\mathcal{V}_{i,1}^{\mathcal N}(1) = \underbrace{(z_i^\star-\hat z_{i,1})}_{\text{attraction}} + \underbrace{(\hat z_{i,1}-\hat z_{i,2})}_{\text{repulsion}}.
\end{equation}
Thus, the unguided field attracts each sample toward its paired target while repelling the two generated samples. For $\alpha_i>1$, mismatched negatives enter with relative weight
$w_i k_{ij}/k_i^{\mathrm{gen}}$, with their locality controlled by the guidance-dependent temperature in Eq.~\eqref{eq:tau_schedule}. Since all samples entering Eq.~\eqref{eq:null_drift_field} lie in
$\mathcal{N}(\mathcal{A}_i)$, the resulting drift field also belongs to this nullspace. Hence:
\begin{equation}
\label{eq:null_drift_consistency}
\mathcal{A}_i\mathcal{V}_{i,r}^{\mathcal N}(\alpha_i)=0, \qquad \mathcal{A}_i \left[ \hat x_{i,r} + \eta\mathcal{V}_{i,r}^{\mathcal N}(\alpha_i) \right]
= y_i 
\end{equation}
for $r\in\{1,2\}$ and any drift step $\eta$.

The generator is trained toward this field using the stop-gradient objective:
\begin{equation}
\label{eq:drift_loss}
\mathcal{L}_{\mathrm{drift}}^{\mathcal N}=\frac{1}{2(3L+1)}\sum_{b=1}^{3L+1}\sum_{r=1}^{2}\left\|\mathcal{W}_b^{(L)}\hat z_{i,r}-\operatorname{sg}\!\left(\mathcal{W}_b^{(L)}\hat z_{i,r}+\mathcal{V}_{i,r,b}^{\mathcal N}\right)\right\|_2^2,
\end{equation}
where $\mathcal{W}_b^{(L)}$ extracts the $b$th subband of an $L$-level orthonormal Haar decomposition, and setting $L=0$ recovers image-domain drift supervision. We additionally use direct image and nullspace supervision:
\begin{align}
\mathcal{L}_{\mathrm{img}} &= \sum_{r=1}^{2} \left\| h_{i,r}-x_i^\star \right\|_2^2, \,\, \mathcal{L}_{\mathrm{null}}= \sum_{r=1}^{2} \left\| \hat z_{i,r}-z_i^\star \right\|_1.
\end{align}
The image loss is assigned a small fixed weight, whereas the nullspace loss provides direct supervision early in training and is cosine annealed over $T$ iterations. Overall, the complete training objective is:
\begin{equation}
\label{eq:driftrecon_loss}
\mathcal{L}_{\mathrm{DriftRecon}} = \lambda_{\mathrm{drift}}\mathcal{L}_{\mathrm{drift}}^{\mathcal N} + \lambda_{\mathrm{img}}\mathcal{L}_{\mathrm{img}}
+ \lambda_{\mathrm{null}}\mathcal{L}_{\mathrm{null}}.
\end{equation}

\begin{table*}[t]
\centering
\caption{Sparse-view CT reconstruction performance at $R=6,\,12$. PSNR (dB), SSIM (\%), and LPIPS ($\times$100) are listed as mean$\pm$std across the test set. Avg. denotes the arithmetic mean across the two acceleration levels. The top-performing method is shown in bold font.}
\label{tab:sparse_view_ct_results}
\setlength{\tabcolsep}{3.0pt}
\resizebox{0.97\textwidth}{!}{%
\begin{tabular}{lcccc|cccc|cccc}
\hline
& \multicolumn{4}{c|}{$R=6$}
& \multicolumn{4}{c|}{$R=12$}
& \multicolumn{4}{c}{Avg.} \\
Method
& PSNR$\uparrow$
& SSIM$\uparrow$
& LPIPS$\downarrow$
& FID$\downarrow$
& PSNR$\uparrow$
& SSIM$\uparrow$
& LPIPS$\downarrow$
& FID$\downarrow$
& PSNR$\uparrow$
& SSIM$\uparrow$
& LPIPS$\downarrow$
& FID$\downarrow$ \\
\hline
FBPConvNet {\scriptsize (Jin et al., 2017)}
& 38.80$\pm$0.65 & 94.93$\pm$0.78 & 3.79$\pm$0.99 & 83.70
& 34.51$\pm$0.80 & 90.00$\pm$1.41 & 7.33$\pm$1.44 & 136.91
& 36.66 & 92.47 & 5.56 & 110.31 \\
MoDL {\scriptsize (Aggarwal et al., 2019)}
& 40.26$\pm$1.12 & 96.18$\pm$0.84 & 2.63$\pm$0.56 & 47.07
& 36.87$\pm$1.21 & 92.67$\pm$1.39 & 4.68$\pm$0.76 & 96.18
& 38.56 & 94.43 & 3.65 & 71.62 \\
MambaRoll {\scriptsize (Kabas et al., 2026)}
& 41.36$\pm$1.14 & 96.92$\pm$0.80 & 1.95$\pm$0.69 & 51.45
& 37.27$\pm$1.12 & 92.95$\pm$1.32 & 4.59$\pm$0.83 & 100.72
& 39.31 & 94.94 & 3.27 & 76.09 \\
RAM {\scriptsize (Terris et al., 2026)}
& 39.99$\pm$0.98 & 96.41$\pm$0.77 & 2.53$\pm$0.78 & 67.31
& 35.91$\pm$0.96 & 92.25$\pm$1.27 & 5.01$\pm$1.06 & 101.67
& 37.95 & 94.33 & 3.77 & 84.49 \\
SSDiffRecon {\scriptsize (Korkmaz et al., 2023)}
& 39.18$\pm$0.81 & 94.49$\pm$0.81 & 4.27$\pm$0.58 & 90.64
& 33.75$\pm$0.87 & 87.98$\pm$1.92 & 9.33$\pm$1.55 & 171.63
& 36.47 & 91.23 & 6.80 & 131.13 \\
FlowDPS {\scriptsize (Kim et al., 2025)}
& 40.37$\pm$1.12 & 96.60$\pm$0.79 & 2.67$\pm$0.54 & 40.24
& 36.66$\pm$1.06 & 93.37$\pm$1.25 & 4.54$\pm$0.64 & 81.98
& 38.52 & 94.98 & 3.60 & 61.11 \\
RDDM {\scriptsize (Wang et al., 2026)}
& 40.88$\pm$1.04 & 96.29$\pm$0.71 & 1.71$\pm$0.44 & 34.97
& 37.15$\pm$1.04 & 92.51$\pm$1.15 & 4.04$\pm$0.62 & 79.27
& 39.02 & 94.40 & 2.88 & 57.12 \\
SA-RDM {\scriptsize (Lyu et al., 2026)}
& 40.95$\pm$1.10 & 96.84$\pm$0.78 & 1.88$\pm$0.66 & 42.66
& 36.87$\pm$1.06 & 92.92$\pm$1.18 & 3.91$\pm$0.77 & 76.12
& 38.91 & 94.88 & 2.90 & 59.39 \\
\hline
DriftRecon {\scriptsize ($\alpha=1$)}
& \textbf{42.06$\pm$1.15} & \textbf{97.45$\pm$0.74}
& \textbf{1.05$\pm$0.41} & \textbf{20.27}
& \textbf{37.59$\pm$1.07} & \textbf{94.02$\pm$1.12}
& \textbf{2.86$\pm$0.55} & \textbf{54.62}
& \textbf{39.83} & \textbf{95.73}
& \textbf{1.96} & \textbf{37.45} \\
\hline
\end{tabular}%
}
\end{table*}

\begin{figure*}[t]
    \centerline{\includegraphics[width=\textwidth]{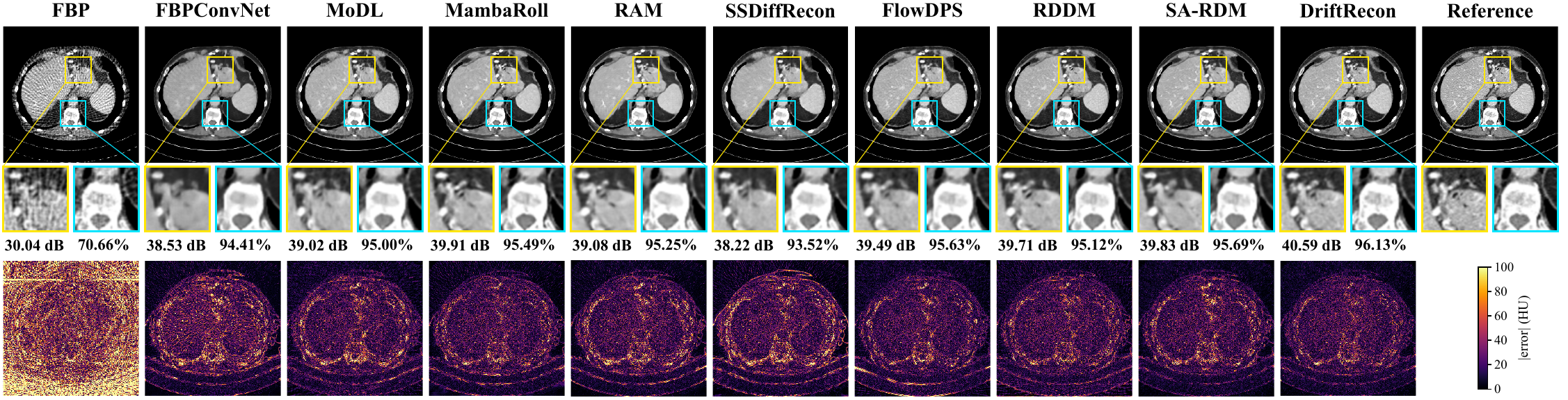}}
    \caption{\textbf{CT reconstruction.} Representative results from an abdominal cross-section at $R=6$. Reconstructions from competing methods are shown together with the sparse-view FBP reconstruction and the full-dose reference image. Images are displayed with an abdominal window of $[-160, 240]$~HU (width 400~HU, level 40~HU). PSNR and SSIM values for each method are marked below the corresponding image. Absolute error maps are displayed in HU over $[0, 100]$~HU (see colorbar).}
    \label{fig:ct}
\end{figure*}

\section{Methods}
\subsection{Datasets}
CT experiments used 1-mm abdominal data from NIH--AAPM--Mayo Clinic Low Dose CT Grand Challenge \cite{mccollough2017lowdose}. Full-dose images were clipped to $[-1000,2000]$~HU, scaled to $[0,1]$, and analyzed at $256\times256$ grid size. Training, validation, and test splits comprised 2,402, 305, and 263 cross-sections without subject overlap. Sparse-view sinograms were generated using a parallel-beam Radon transform with 364 detector bins and uniformly undersampled along the angular dimension at $R=6,12$.

MRI experiments used multi-contrast brain MRI from M4Raw \cite{lyu2023m4raw}, acquired at 0.3~T with a four-channel coil. T\textsubscript{1}-weighted, T\textsubscript{2}-weighted, and FLAIR acquisitions were scaled to [0, 1] via maximum-intensity normalization and analyzed at $256\times256$ grid size, with coil sensitivities estimated using ESPIRiT \cite{uecker2014espirit}. Training, validation, and test splits comprised 18,432, 4,320, and 7,200 cross-sections without subject overlap. Data were undersampled at $R=4,8$ using 2D Gaussian variable-density patterns \cite{lustig2007sparse}.

\subsection{Implementation Details}
DriftRecon used a four-level multiscale convolutional generator with feature dimensionalities of 64, 128, 256, and 512 and four residual blocks per stage \cite{he2016deep}. Measurement conditioning was incorporated within each residual block as in \cite{terris2026ram}, and $\alpha$ was provided through a learned embedding. For CT, drift supervision used $L=3$, $M=2$ similarity kernels with $\tau_{m,0}=\{0.2,1.0\}$, and $\lambda_{\mathrm{img}}=1$. For MRI, drift supervision used $L=0$, $M=1$ with $\tau_{1,0}=1.0$, and $\lambda_{\mathrm{img}}=0.01$. For both modalities, $\sigma=0.05$, $\beta=0.5$, and $\lambda_{\mathrm{drift}}=1$, while $\lambda_{\mathrm{null}}$ was cosine annealed from $0.01$ to $0$. Guidance used $\alpha=1$ with probability $0.5$ and otherwise $\alpha\sim p(\alpha)\propto\alpha^{-3}$ on $[1,4]$.

\begin{table*}[t]
\centering
\caption{Accelerated MRI reconstruction performance at $R=4,\,8$. PSNR (dB), SSIM (\%), and LPIPS ($\times$100) are listed as mean$\pm$std across the test set. Avg. denotes mean metrics across $R$. The top-performing method is shown in bold font.}
\label{tab:accelerated_mri_results}
\setlength{\tabcolsep}{3.0pt}
\resizebox{0.97\textwidth}{!}{%
\begin{tabular}{lcccc|cccc|cccc}
\hline
& \multicolumn{4}{c|}{$R=4$}
& \multicolumn{4}{c|}{$R=8$}
& \multicolumn{4}{c}{Avg.} \\
Method
& PSNR$\uparrow$
& SSIM$\uparrow$
& LPIPS$\downarrow$
& FID$\downarrow$
& PSNR$\uparrow$
& SSIM$\uparrow$
& LPIPS$\downarrow$
& FID$\downarrow$
& PSNR$\uparrow$
& SSIM$\uparrow$
& LPIPS$\downarrow$
& FID$\downarrow$ \\
\hline
UNet {\scriptsize (Muckley et al., 2021)}
& 29.15$\pm$4.24 & 78.39$\pm$5.05 & 12.70$\pm$9.79 & 24.05
& 25.46$\pm$7.89 & 72.08$\pm$6.65 & 24.15$\pm$17.41 & 34.74
& 27.31 & 75.24 & 18.43 & 29.40 \\
MoDL {\scriptsize (Aggarwal et al., 2019)}
& 31.22$\pm$2.08 & 81.67$\pm$4.90 & 8.50$\pm$1.46 & 16.83
& 29.71$\pm$2.09 & 76.61$\pm$6.27 & 14.40$\pm$2.36 & 24.43
& 30.47 & 79.14 & 11.45 & 20.63 \\
MambaRoll {\scriptsize (Kabas et al., 2026)}
& 31.53$\pm$2.06 & 82.79$\pm$4.59 & 7.54$\pm$1.58 & 14.20
& 30.02$\pm$2.11 & 77.50$\pm$5.60 & 13.70$\pm$2.60 & 20.18
& 30.78 & 80.15 & 10.62 & 17.19 \\
RAM {\scriptsize (Terris et al., 2026)}
& 31.68$\pm$2.10 & 83.10$\pm$4.58 & 7.27$\pm$1.53 & 13.10
& 30.21$\pm$2.15 & 78.00$\pm$5.57 & 13.06$\pm$2.48 & 24.07
& 30.95 & 80.55 & 10.17 & 18.59 \\
SSDiffRecon {\scriptsize (Korkmaz et al., 2023)}
& 31.27$\pm$1.85 & 82.63$\pm$4.73 & 5.78$\pm$1.12 & 10.94
& 29.58$\pm$1.79 & 77.48$\pm$5.78 & 11.48$\pm$1.87 & 22.27
& 30.43 & 80.06 & 8.63 & 16.61 \\
FlowDPS {\scriptsize (Kim et al., 2025)}
& 31.47$\pm$1.90 & 82.35$\pm$5.12 & 6.68$\pm$1.48 & 9.26
& 30.05$\pm$1.88 & 78.39$\pm$5.81 & 9.75$\pm$2.07 & 14.33
& 30.76 & 80.37 & 8.22 & 11.80 \\
RDDM {\scriptsize (Wang et al., 2026)}
& 29.68$\pm$1.55 & 79.26$\pm$5.42 & 6.59$\pm$1.79 & 16.93
& 28.60$\pm$1.57 & 75.47$\pm$6.31 & 11.51$\pm$2.69 & 25.54
& 29.14 & 77.37 & 9.05 & 21.24 \\
SA-RDM {\scriptsize (Lyu et al., 2026)}
& 31.57$\pm$1.92 & 82.59$\pm$5.34 & 5.98$\pm$1.58 & 9.06
& 30.31$\pm$1.94 & 78.72$\pm$6.36 & 10.45$\pm$2.30 & 16.64
& 30.94 & 80.66 & 8.22 & 12.85 \\
\hline
DriftRecon ($\alpha=1$)
& \textbf{32.16$\pm$2.02} & \textbf{84.25$\pm$4.62}
& \textbf{4.20$\pm$1.40} & \textbf{5.88}
& \textbf{30.58$\pm$2.12} & \textbf{79.86$\pm$5.63}
& \textbf{9.02$\pm$2.31} & \textbf{10.51}
& \textbf{31.37} & \textbf{82.06}
& \textbf{6.61} & \textbf{8.20} \\
\hline
\end{tabular}%
}
\end{table*}
\begin{figure*}[t]
    \centerline{\includegraphics[width=\textwidth]{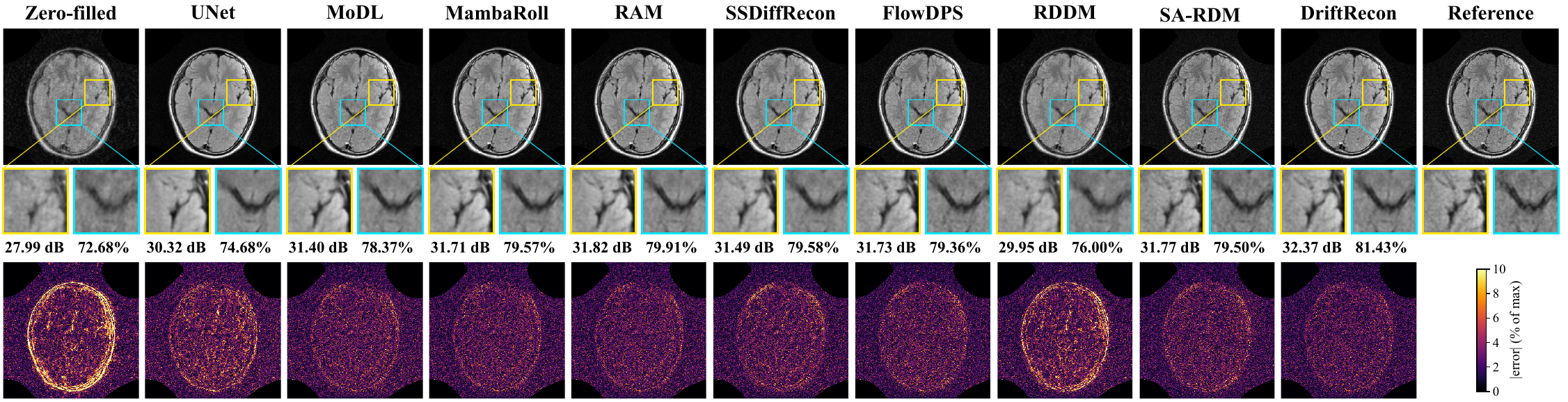}}
    \caption{\textbf{Accelerated MRI reconstruction.} Representative results from a FLAIR cross-section at $R=4$. Reconstructions from competing methods are shown together with the traditional zero-filled reconstruction and the reference image from a fully-sampled acquisition. Absolute error maps relative to the reference image are displayed as a percentage of the maximum reference intensity (see colorbar).}
    \label{fig:mri}
\end{figure*}

\subsection{Competing Methods}
Baselines included modality-specific data-driven methods (UNet \cite{muckley2021fastmri} for MRI; FBPConvNet \cite{unser2017} for CT), physics-driven methods (MoDL \cite{MoDl}, MambaRoll \cite{kabas2026mambaroll}, RAM \cite{terris2026ram}), few-step generative methods (SSDiffRecon \cite{ssdiffrecon} and FlowDPS \cite{kim2025flowdps}, each with five sampling steps), and one-step drifting methods (RDDM \cite{wang2026rddm}, SA-RDM \cite{lyu2026sardmdc}). All methods operated on individual cross-sections, with MRI represented using two channels for the real and imaginary components and CT using a single channel. All models were trained from random initialization using the same training data and imaging operators. Competing methods followed their published architectures and objectives, with key hyperparameters selected on the validation sets. Physics-driven unrolled methods used $K=3$ cascades for CT and $K=5$ for MRI. Models were trained with Adam for 80k iterations for CT and 20k iterations for MRI, with the learning rates selected from $10^{-6}$--$10^{-3}$ and halved every 10k iterations. Performance was assessed using PSNR, SSIM, LPIPS, and FID against reference CT or MR images from fully-sampled acquisitions. Performance differences were assessed using Wilcoxon signed-rank tests.

\section{Results}

\subsection{Sparse-View CT Reconstruction}
\label{sec:comparison_ct}

Table~\ref{tab:sparse_view_ct_results} lists performance in CT reconstruction at $R=6,12$. For analysis, baselines are grouped into physics-driven methods (MoDL, MambaRoll, RAM), few-step generative methods (SSDiffRecon, FlowDPS), and drifting methods (RDDM, SA-RDM). DriftRecon achieves the best PSNR, SSIM, LPIPS, and FID at both acceleration levels (p$<$0.05). On average, DriftRecon yields improvements over physics-driven baselines of 1.2dB PSNR, 1.2\% SSIM, 1.6 LPIPS, and 40.0 FID. Against few-step generative baselines, the corresponding improvements are 2.3dB PSNR, 2.6\% SSIM, 3.2 LPIPS, and 58.7 FID. Relative to drifting baselines, the improvements are 0.9dB PSNR, 1.1\% SSIM, 0.9 LPIPS, and 20.8 FID. The gains in LPIPS and FID are accompanied by the highest PSNR and SSIM, indicating that DriftRecon improves perceptual similarity and distributional agreement without sacrificing reference fidelity.

Representative reconstructions at $R=6$ are displayed in Fig.~\ref{fig:ct}. SSDiffRecon exhibits visible smoothing, while FlowDPS better preserves local contrast but yields residual errors around tissue interfaces. Among the physics-driven methods, MoDL and RAM show more spatially distributed residual errors, whereas MambaRoll provides sharper tissue depiction but retains localized discrepancies along organ boundaries. The drifting baselines preserve local structure more effectively, although residual errors remain around hepatic vessels and peripheral tissue interfaces. DriftRecon most closely reproduces the reference, preserving finer intrahepatic structure while reducing residual errors across the liver parenchyma and surrounding anatomy. The visual comparison is consistent with the quantitative gains, showing improved preservation of fine anatomical detail together with low reconstruction error.

\begin{table*}[t]
\centering
\setlength{\tabcolsep}{2.5pt}
\begin{minipage}[t]{0.24\textwidth}\vspace{0pt}
\caption{Effect of guidance strength under conventional classifier-free guidance (CFG) and guidance-coupled temperature scheduling. Results are shown for $\alpha=1$--$4$, together with the metric differences between $\alpha=4$ and $\alpha=1$.}
\label{tab:guidance_sweep}
\end{minipage}
\hfill
\begin{minipage}[t]{0.73\textwidth}\vspace{0pt}
\resizebox{0.9\linewidth}{!}{%
\begin{tabular}{lc|cccc|cccc}
\hline
& &
\multicolumn{4}{c|}{CT} &
\multicolumn{4}{c}{MRI} \\
\cline{3-6}\cline{7-10}
Method & $\alpha$
& PSNR $\uparrow$
& SSIM $\uparrow$
& LPIPS $\downarrow$
& FID $\downarrow$
& PSNR $\uparrow$
& SSIM $\uparrow$
& LPIPS $\downarrow$
& FID $\downarrow$ \\
\hline
Conv. CFG
& 1
& 41.99$\pm$1.15 & 97.45$\pm$0.73 & 1.12$\pm$0.42 & 21.72
& 32.14$\pm$2.03 & 84.17$\pm$4.58 & 4.37$\pm$1.44 & 6.69 \\
& 2
& 41.99$\pm$1.16 & 97.43$\pm$0.73 & 1.02$\pm$0.40 & 21.24
& 31.71$\pm$1.91 & 83.89$\pm$4.63 & 4.78$\pm$1.65 & 8.03 \\
& 3
& 41.20$\pm$1.11 & 97.27$\pm$0.74 & 1.04$\pm$0.39 & 21.27
& 30.95$\pm$1.67 & 82.99$\pm$4.62 & 5.20$\pm$1.79 & 10.39 \\
& 4
& 39.88$\pm$1.03 & 96.97$\pm$0.75 & 1.19$\pm$0.40 & 22.79
& 30.27$\pm$1.49 & 81.92$\pm$4.57 & 5.68$\pm$1.85 & 13.55 \\
& $\Delta_{1\rightarrow4}$
& -2.11 & -0.48 & +0.07 & +1.07
& -1.87 & -2.25 & +1.31 & +6.86 \\
\hline
DriftRecon
& 1
& 42.01$\pm$1.15 & 97.42$\pm$0.73 & 1.09$\pm$0.40 & 21.34
& 32.15$\pm$2.02 & 84.20$\pm$4.57 & 4.32$\pm$1.42 & 6.61 \\
& 2
& 41.91$\pm$1.15 & 97.39$\pm$0.73 & 0.99$\pm$0.38 & 20.72
& 31.85$\pm$1.94 & 84.01$\pm$4.60 & 4.66$\pm$1.60 & 7.64 \\
& 3
& 41.58$\pm$1.13 & 97.32$\pm$0.74 & 0.97$\pm$0.37 & 20.54
& 31.57$\pm$1.85 & 83.72$\pm$4.61 & 4.85$\pm$1.69 & 8.44 \\
& 4
& 41.25$\pm$1.11 & 97.25$\pm$0.74 & 0.97$\pm$0.37 & 20.52
& 31.45$\pm$1.82 & 83.57$\pm$4.61 & 4.94$\pm$1.72 & 8.85 \\
& $\Delta_{1\rightarrow4}$
& \textbf{-0.76} & \textbf{-0.17} & \textbf{-0.12} & \textbf{-0.82}
& \textbf{-0.70} & \textbf{-0.63} & \textbf{+0.62} & \textbf{+2.24} \\
\hline
\end{tabular}}
\end{minipage}
\end{table*}

\subsection{Accelerated MRI Reconstruction}
\label{sec:comparison_mri}

Table~\ref{tab:accelerated_mri_results} lists performance in MRI reconstruction at $R=4,8$. Again, DriftRecon achieves the best performance metrics across acceleration levels (p$<$0.05). On average, it yields improvements over physics-driven baselines of 0.6dB PSNR, 2.1\% SSIM, 4.1 LPIPS, and 10.6 FID. Against few-step generative baselines, the improvements are 0.8dB PSNR, 1.8\% SSIM, 1.8 LPIPS, and 6.0 FID. Relative to drifting baselines, the improvements are 1.3dB PSNR, 3.0\% SSIM, 2.0 LPIPS, and 8.8 FID. Thus, DriftRecon maintains the strongest reference fidelity in PSNR and SSIM while also improving perceptual similarity and distributional agreement.

Representative FLAIR reconstructions at $R=4$ are shown in Fig.~\ref{fig:mri}. UNet exhibits visible smoothing of tissue contrast and residual errors along the brain boundary, while SSDiffRecon and MoDL show milder blurring of internal structures. FlowDPS, MambaRoll, and RAM better preserve anatomical detail, but retain spatially distributed residual errors across the brain parenchyma. RDDM shows the most pronounced residual boundary errors among the generative methods, together with reduced tissue definition, whereas SA-RDM provides improved structural recovery but retains residual errors in internal tissue contrast. DriftRecon most closely reproduces the reference FLAIR contrast and anatomical detail, with reduced residual errors throughout the brain tissue. This visual behavior is consistent with its stronger reference fidelity and perceptual similarity.

\subsection{Effect of Guidance-Coupled Temperature Scheduling}
Increasing the guidance strength $\alpha$ places greater emphasis on image characteristics learned by the prior, including high-spatial-frequency detail that is weakly constrained by undersampled measurements. This shifts the reconstruction from subject-specific anatomy supported by the acquired data toward characteristics represented more strongly by the learned prior. Under conventional guidance with a fixed similarity-kernel temperature (CFG), stronger guidance can therefore introduce plausible but subject-inaccurate detail, alter local contrast, or amplify noise and artifacts. These changes can reduce agreement with the reference images, degrading subject-level PSNR, SSIM, and LPIPS as well as distribution-level FID. The proposed guidance-coupled temperature scheduling alleviates these losses by progressively localizing the drift interactions strengthened at higher $\alpha$, limiting guidance-induced deviations while retaining prior-supported image detail.

For a controlled comparison, CFG was implemented within the same DriftRecon formulation by setting $\beta=0$, which keeps the similarity-kernel temperature fixed, whereas guidance-coupled scheduling used $\beta=0.25$. Table~\ref{tab:guidance_sweep} compares the two formulations over $\alpha=1$--$4$ and reports the metric changes between $\alpha=1$ and $\alpha=4$. Compared with CFG, DriftRecon improves the $\Delta_{1\rightarrow4}$ changes by 1.4dB PSNR, 0.3\% SSIM, 0.2 LPIPS, and 1.9 FID in CT. The corresponding improvements in MRI are 1.2dB PSNR, 1.6\% SSIM, 0.7 LPIPS, and 4.6 FID. Thus, coupling kernel temperature to guidance strength consistently limits the degradation in reference-based quality metrics as guidance is increased.

\begin{figure}[t]
    \centering
    \includegraphics[width=0.9\columnwidth]{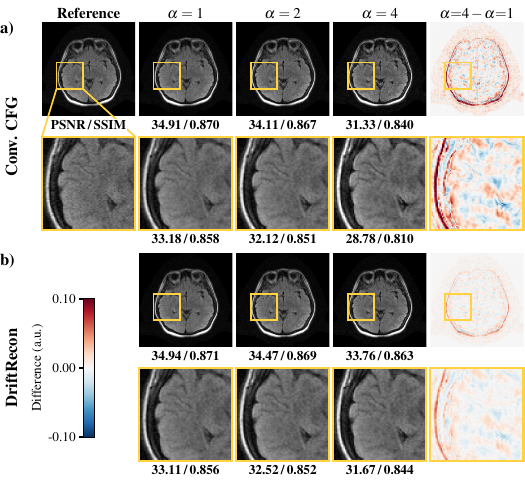}
\caption{\textbf{Effect of guidance strength.} Representative reconstructions for \textbf{(a)} conventional classifier-free guidance (CFG) with fixed similarity-kernel temperature, and \textbf{(b)} DriftRecon with guidance-coupled temperature scheduling at $\alpha=1,2,4$. Enlarged regions and difference maps between $\alpha=4$ and $\alpha=1$ are shown.}
    \label{fig:alpha_sweep}
\end{figure}

Fig.~\ref{fig:alpha_sweep} provides a representative visual comparison of this behavior. Under CFG, increasing guidance produces progressively larger changes in tissue structure and local contrast, whereas DriftRecon better preserves subject-specific image content while still allowing sharper local tissue depiction. The enlarged region further shows that the proposed schedule retains finer structural detail with substantially smaller deviations from the $\alpha=1$ reconstruction.

\subsection{Computational Complexity}
Table~\ref{tab:complexity} compares inference time and memory requirements across reconstruction methods. For CT, DriftRecon requires 23.56~ms and 0.74~GB per cross-section, substantially lower than the few-step generative baselines in both inference time and memory. For MRI, its inference time of 68.39~ms and memory requirement of 0.40~GB remain within the range of the few-step generative baselines. Computational cost relative to the physics-driven and one-step drifting baselines is more modality dependent, reflecting differences in model architecture and inference strategy. Across both modalities, DriftRecon maintains inference memory below 0.75~GB while producing each reconstruction in a single forward pass without iterative generative sampling.

\subsection{Ablation Studies}
Table~\ref{tab:ablation} examines the performance contributions from the main design elements of DriftRecon. The `w/o drifting' variant uses the same multiscale generator trained to minimize $\ell_2$-loss in image domain without drift supervision. The `w/o nullspace' variant performs drifting in image space without nullspace projection, followed by hard data consistency at the generator output. The `w/o meas. cond.' variant removes measurement conditioning throughout the generator. The `w/o aux. losses' variant retains only the drift loss during training. Finally, the `w/o wavelet' variant adopts image-domain in place of multiscale Haar-domain drift supervision. The complete formulation provides the strongest results across both modalities, with particularly large degradations when drifting or measurement conditioning is removed, indicating that all method components contribute to reconstruction performance.

\begin{table}[t]
\centering
\caption{Inference time (Inf.; ms), memory (Mem.; GB), and GFLOPs reported per cross-section.}
\label{tab:complexity}
\setlength{\tabcolsep}{2.0pt}
\resizebox{0.9\columnwidth}{!}{%
\begin{tabular}{lccc|ccc}
\hline
& \multicolumn{3}{c|}{CT}
& \multicolumn{3}{c}{MRI} \\
\cline{2-7}
Method
& Inf. & Mem. & GFLOP
& Inf. & Mem. & GFLOP \\
\hline
FBPConvNet/UNet
& 2.11 & 0.25 & 101.0
& 2.11 & 0.28 & 101.1 \\
SSDiffRecon
& 367.01 & 0.87 & 121.3
& 39.03 & 0.28 & 205.0 \\
FlowDPS
& 73.79 & 1.39 & 2769.1
& 70.49 & 1.02 & 2770.6 \\
MoDL
& 6.67 & 0.45 & 43.9
& 2.62 & 0.09 & 74.0 \\
MambaRoll
& 106.18 & 0.83 & 31.6
& 7.74 & 0.09 & 56.3 \\
RAM
& 32.80 & 0.70 & 291.8
& 36.11 & 0.34 & 308.3 \\
RDDM
& 5.63 & 0.27 & 123.4
& 6.91 & 0.26 & 123.6 \\
SA-RDM
& 21.72 & 1.94 & 246.8
& 14.56 & 0.45 & 248.0 \\
DriftRecon
& 23.56 & 0.74 & 417.9
& 68.39 & 0.40 & 308.1 \\
\hline
\end{tabular}%
}
\end{table}

\section{Discussion}
DriftRecon combines one-step generative reconstruction with explicit constraints from the imaging operator. Operator-nullspace drifting confines the learned prior to image components unresolved by the acquired measurements, while measurement conditioning incorporates the forward model throughout the generator. Guidance-coupled temperature scheduling provides continuous control over the fidelity--realism trade-off by progressively localizing drift interactions as guidance increases. Across accelerated MRI and sparse-view CT, DriftRecon outperformed established physics-driven and generative reconstruction methods as well as recent drifting approaches while requiring a single generator evaluation. Ablation studies supported the contributions of operator-nullspace drifting, measurement conditioning, and auxiliary supervision, while the guidance analysis showed that the proposed temperature schedule substantially limits fidelity loss at increased guidance strengths.

The present formulation transfers progressive distribution refinement from inference to training and reconstructs each image in a single generator evaluation. Additional inference computation could nevertheless be useful when further refinement is warranted. Drift Flow Matching recently combined direct drifting with optional flow-based refinement, enabling test-time scaling between one-step generation and iterative transport \cite{driftflowmatching}. Diffusion-bridge formulations have also been investigated for medical image synthesis \cite{SelfRDB}. A related extension of DriftRecon could retain one-step reconstruction as the default while allowing a small number of nullspace-restricted refinement steps for challenging acquisitions or when additional computation is available, without returning to the long inference of conventional diffusion models.

\begin{table}[t]
\centering
\setlength{\tabcolsep}{2pt}
\caption{Ablation of method components for sparse-view CT at $R=6$ and accelerated MRI at $R=4$.}
\label{tab:ablation}
\resizebox{0.99\linewidth}{!}{%
\begin{tabular}{lcccc|cccc}
\hline
& \multicolumn{4}{c}{CT}
& \multicolumn{4}{c}{MRI} \\
\cline{2-9}
Variant
& PSNR & SSIM & LPIPS & FID
& PSNR & SSIM & LPIPS & FID \\
\hline
w/o drifting
& 39.99 & 96.41 & 2.53 & 67.31
& 31.68 & 83.10 & 7.27 & 13.10 \\
w/o nullspace
& 41.91 & 97.37 & 1.19 & 22.20
& 32.01 & 83.70 & 4.31 & 6.99 \\
w/o meas. cond.
& 38.02 & 95.35 & 3.71 & 58.21
& 31.70 & 82.34 & 6.64 & 14.79 \\
w/o aux. losses
& 41.30 & 96.96 & 1.24 & 21.53
& 32.06 & 82.47 & 4.43 & 7.23 \\
w/o wavelet
& 41.21 & 96.93 & 1.09 & 20.90
& 32.15 & 84.18 & 4.35 & 6.83 \\
\hline
DriftRecon
& 42.06 & 97.45 & 1.05 & 20.27
& 32.16 & 84.25 & 4.20 & 5.88 \\
\hline
\end{tabular}%
}
\end{table}

A second direction concerns inference-time adaptation of the fidelity--realism trade-off. DriftRecon currently uses a global guidance strength together with the corresponding guidance-dependent temperature. These quantities could instead be determined from the measurements or current reconstruction, allowing guidance to adapt to acceleration level, sampling pattern, or measurement quality. Acquisition-adaptive reconstruction has previously been considered in model-based networks \cite{pramanik2023adamodl}, and related adaptation may be useful when acquisition characteristics vary across sites. Federated reconstruction provides another relevant setting, particularly when institutions differ in sampling protocols, data distributions, or reconstruction models \cite{Nezhad2025GenerativeReconstruction}. Extending DriftRecon in this direction could permit acquisition- or site-specific guidance while retaining a common generative formulation.

The operator-nullspace formulation is independent of the particular reconstruction module used within the generator. State-space models have recently been investigated for physics-guided MRI reconstruction \cite{mambarecon}, while MambaRoll has demonstrated multiscale state-space modeling for accelerated MRI and sparse-view CT reconstruction \cite{kabas2026mambaroll}. Such modules could be incorporated into DriftRecon while retaining the same restriction of the generated component to the operator nullspace. State-space modeling has also been extended to multimodal medical image synthesis \cite{I2IMamba}, providing a possible direction for extending drifting beyond reconstruction to conditional image synthesis.

The accuracy of the nullspace decomposition depends on the forward model used during reconstruction. In MRI, errors in coil-sensitivity estimation or trajectory specification alter the encoding operator, while CT reconstruction can be affected by inaccuracies in acquisition geometry or system calibration. Future work can examine robustness to such model mismatch and, where needed, jointly refine the relevant acquisition parameters during reconstruction. Extension to non-Cartesian MRI and additional CT acquisition geometries would likewise require appropriate realization of the corresponding imaging operator and nullspace. These factors determine how accurately measurement-supported and unresolved image components can be separated in practice.

\section{Conclusion}
We introduced DriftRecon, a one-step medical image reconstruction method that models distributional drift directly in the nullspace of the imaging operator. DriftRecon combines nullspace-restricted perturbations and generated components with guidance-coupled temperature scheduling to preserve measurement-supported image content while controlling the fidelity--realism trade-off. Across accelerated MRI and sparse-view CT, DriftRecon outperformed established generative reconstruction methods and recent drifting approaches while requiring a single generator evaluation. These results support operator-nullspace drifting as an effective formulation for efficient generative medical image reconstruction.

\bibliographystyle{IEEEtran} 
\bibliography{IEEEabrv,refs}

\end{document}